# Campaigning through the lens of Google: A large-scale algorithm audit of Google searches in the run-up to the Swiss Federal Elections 2023

Tobias Rohrbach[1]*, tobias.rohrbach@unibe.ch

Mykola Makhortykh[1], mykola.makhortykh@unibe.ch

Maryna Sydorova, maryna.sydorova@unibe.ch

[1]*Institute of Communication and Media Studies, University of Bern, Bern, Switzerland*

*\*Corresponding author*

**Abstract**: Search engines like Google have become major sources of information for voters during election campaigns. To assess potential biases across candidates' gender and partisan identities in the algorithmic curation of candidate information, we conducted a large-scale algorithm audit analyzing Google's selection and ranking of information about candidates for the 2023 Swiss Federal Elections, three and one week before the election day. Results indicate that text searches prioritize media sources in search output but less so for women politicians. Image searches revealed a tendency to reinforce stereotypes about women candidates, marked by a disproportionate focus on stereotypically pleasant emotions for women, particularly among right-leaning candidates. Crucially, we find that patterns of candidates' representation in Google text and image searches are predictive of their electoral performance.

**Funding statement**

This research was enabled through funding by the Swiss Federal Office of Communications awarded to the authors of this manuscript.

**Campaigning through the lens of Google: A large-scale algorithm audit of Google searches in the run-up to the Swiss Federal Elections 2023**

Search engines like Google or Bing are used by individuals not only to find practical information, such as the address of the closest post office or a pizzeria, but also to stay informed about important societal developments, such as elections. Trusted as a reliable source of information (Schultheiß & Lewandowski, 2023), search engines have increasingly become a source of news (or starting point thereof), turning them into key information gatekeepers in liberal democracies (Bro & Wallberg, 2014; Wallace, 2018; Evans et al., 2023). Therefore, understanding how search engines shape individual and collective understanding of social and political reality has become a core concern of digital journalism research(Westlund et al., 2023; Pradel, 2021; Haim et al., 2018; Trielli & Diakopoulos, 2022; Urman et al., 2022; Ulloa et al., 2024a). In the case of political elections, the relevance of search engines' performance has been amplified by evidence that the selection of information by search algorithms has an impact on voter preferences (Epstein & Robertson, 2015; Puschmann, 2019). Consequently, the way search engines select and rank information about electoral matters—above all, the candidates running in elections—shapes the core processes of democratic decision-making.

Despite the importance of search engines for the public sphere, their functionalities and effects on the processes of political decision-making remain untransparent. The lack of transparency is particularly concerning due to plentiful evidence of search engines' malperformance, ranging from non-systematic errors to systematic bias. Often understood as a systematic skewness towards specific individuals or groups (Friedman & Nissenbaum, 1996), algorithmic (search) bias can give rise to distorted representations of (vulnerable) groups. In the realm of political communication, multiple studies demonstrate that search engines tend to cast women as less competent than men by associating them with undesirable personality traits, overemphasizing emotional characteristics, showing them with lower face-to-body ratios, or even directly sexualizing their representation (Otterbacher et al., 2017; Urman & Makhortykh, 2024; Guilbeault et al., 2024). In the context of politics-related information searches, such imbalances in search engine behavior can diminish women's electoral success and sustain their underrepresentation in positions of political decision-making (Rohrbach et al., 2024).

In this article, we aim to assess the possible risks associated with the impact of Google searches on democratic decision-making in Switzerland, with a particular emphasis on the possibility of bias in its outputs towards more vulnerable groups of candidates. For this purpose, we conducted a large-scale audit of Google's image and text searches during the 2023 elections in Switzerland and connect search engine output to election data. First, we investigate potential imbalances in search results by asking to what extent textual and visual search results about candidates differ according to their party and gender. Second, we assess the impact of search engines on election outcomes by quantifying the predictive role of key search results measures for candidates' electoral support.

The results show that text searches prioritize news authoritative sources, particularly government websites and news media. Crucially, text search output included more and higher rank links to media sources for queries of women candidates, which was negatively associated with their electoral performance. Images searches indicated a slight underrepresentation of women in image searches and emotional displays that are in line with prescriptive gender stereotypes. However, the results suggest that the electoral impact of gendered patterns may present both an obstacle and potential opportunity for women candidates, highlighting the complex role of search engines in shaping political narratives.

**Search engines, campaigning, and electoral success**

With ongoing digitalization, the study of journalistic gatekeeping has increasingly focused on influences that curate information flows in online environments (Wallace, 2018; Evans et al., 2023; Makhortykh et al., 2025a; Vermeer et al., 2020). Search engines, particularly Google, constitute "important arbiters of political information" (Puschmann, 2019, p. 826) by serving as primary gateways for accessing knowledge about politics. Research indicates that voters increasingly rely on search engines to better understand political matters, making the digital footprint of electoral discourse more complex (Newman et al., 2023). Some evidence even suggests that people prefer algorithmic information curation by search engines over that by journalists (Thurman et al., 2019; Fisher et al., 2019). Beyond merely providing information, existing work suggests that search engines encourage voters to actively construct narratives around political candidates and their associated policies resulting from retrieved information (Trevisan et al., 2023; Pradel, 2021).

Search engines therefore present algorithmic "machine gatekeepers" (Evans et al., 2023, p. 1682) whose influence on information-seeking process follows a similar logic to traditional journalistic selection and editing principles. By highlighting or giving weight to one type of candidate information over another, search engines curate which aspects of candidates' profiles gain visibility and are made available for voters' impression formation. However, this curating role of search engines in shaping electoral narratives is accompanied by a pressing concern regarding potential biases, especially in relation to politicians' gender and political party affiliation. Studies have uncovered systemic biases in the presentation of search results that can disadvantage certain candidates based on their gender or their political alignment. In the following, we separately review key sources of potential bias in Google text and image searches as well as their impact on election outcome.

**Text searches: Algorithmic curation and media source prominence**

Media visibility has long been a critical factor in candidates' electoral success, as mass media serve as the primary channel through which voters access substantive information about candidates (Esser & Strömbäck, 2014). Operating within the logic of institutionalized journalism, traditional media act as gatekeepers, determining whose voices are amplified in political coverage (Wallace, 2018). Content-analytic studies have quantified disparities in visibility across party and gender lines, revealing structural biases: larger and ruling parties tend to dominate electoral reporting (Hopmann et al., 2012), while women politicians tend to be underrepresented across party context. A meta-analysis of 70 studies confirms a persistent gender gap, particularly in proportional electoral systems (Van der Pas & Aaldering, 2022). Crucially, media effects research demonstrates that greater media visibility—whether at the level of parties (Hoppmann et al., 2010; Geers & Bos, 2017) or individual candidates (Aaldering et al., 2018; van Erkel et al., 2020)—correlates with stronger electoral performance.

This media visibility advantage extends to the digital realm, where search engines introduce an additional, algorithmically driven layer of gatekeeping (Wallace, 2018; Evans et al., 2023). Like traditional media, search engines structure the political information environment by selectively indexing and ranking information sources about candidates (Urman et al., 2022; Schwab et al., 2022; Pradel, 2021). Whereas journalists rely on journalistic criteria (such as news values) when making their decisions, less is known about the underlying algorithmic logic driving the curation of specific sources by search engines. One key observation, however, is that search engines favor domains linking to high-authority sources, including legacy media (Haim et al., 2018; Urman et al., 2022; Norocel & Lewandowski, 2023; Makhortykh et al., 2025a). Moreover, evidence from tracking studies suggest that users' online information journeys often start on search engines and end on websites of traditional news outlets (Vermeer et al., 2020; Makhortykh et al., 2025b). One explanation for the high prominence of media sources might be that search engines proclaim favoring reliable information and promote, as a form of quality optimization, sources with professional accountability (Giomelakis & Veglis, 2016). Moreover, media organizations increasingly rely on search engine optimization techniques to favorably impact their position in ranking of search queries (Schwab et al., 2022).

Therefore, search engines can modulate candidates' media visibility in two ways: (1) by including *more* search results that link to media sources about candidates and (2) by ranking media sources *more highly* in the search results. The combination of these two algorithmic mechanisms is what we term media source prominence on search engines. In traditional offline media, media visibility of candidates during campaigns can be directly measured as the volume of political reporting dedicated to candidates' positions, activities, and personality. Likewise, we argue that media source prominence represents an equivalent measure of candidates' media visibility on search engines. As the presence in traditional media is a necessary condition for media source prominence, we expect search engines to reproduce existing offline power imbalances across gender (*H1a*) and party (*H1b*) lines (Puschmann, 2019; Pradel, 2021; Hoppmann et al., 2012). And while on- and offline mechanisms of the visibility advantage may slightly differ, the underlying principle remains: in both traditional and digital media ecosystems, a larger media presence of candidates should translate into electoral advantage (*H2*).

> *H1*: Search results of (a) men candidates and of (b) candidates running for larger political parties have higher prominence of media sources in Google text searches
>
> *H2*: Candidates whose searches have a higher prominence of media sources in text search results receive more electoral support.

**Image searches: Visual stereotyping of political candidates**

The output of image searches is visual information. Research demonstrates that images may be particularly effective at shaping public perceptions of political issues, events, and candidates (e.g., Jungblut & Haim, 2023). Voters often rely on visual candidate cues, especially from faces, to make heuristic judgements about candidates' traits, qualifications, and their own vote decision (Todorov et al., 2005).

Yet visual representations of candidates—online and in traditional media—have been shown to quantitatively and qualitatively emphasize stereotypical gender roles, impacting candidates' visibility and electoral competitiveness (Diakopoulos et al., 2018; Rohrbach et al., 2023; Jungblut & Haim, 2023). From a quantitative perspective, such gender-stereotypical representations cast women as a statistical minority in a masculinized political sphere. For example, studies on the representation of professional occupations in Google image searches suggest that underrepresentation of women was particularly likely to prime and amplify gender bias and thus "come at a critical social cost" (Guilbeault et al., 2024, p. 6; Otterbacher et al., 2017). A recent large-scale audit of Google image searches for political queries in 60 countries found that the share of women's representation in search output was roughly 29%, and that algorithmically driven underrepresentation of women makes a substantive contribution to constructing masculinized images of politics (Rohrbach et al., 2024). In this sense, the extent of women's presence in image searches should act as a proxy that signals distance to the masculinized center of political power.

In more qualitative terms of image content, patterns of gender-stereotypical representation of women politicians include linking them to communal attributes—that is, traits emphasizing warmth, cooperation, compassion, and harmony. Such gender-stereotypical associations present an electoral liability as they are incongruent with the more assertive traits expected in leaders (Rohrbach, 2024; Bauer & Carpinella, 2018). A key means of visually conveying communality is through emotional displays, specifically through a disproportionate focus on pleasant and positive emotions for women candidates, with images depicting happy, calm, and smiling candidates (Brescoll, 2016). By the same gendered standards, women ought to refrain from expressing negativity, such as anger or disgust, as this might prompt voter backlash in response to violating the communality expectation (Bast, 2024). The idea that search engines mirror—and even exacerbate—prescriptive gendered patterns in emotional displays is evidenced by a recent large-scale analysis of Bing image searches of all candidates running for the 2019 European Election (Jungblut & Haim, 2023). The study found that images of women politicians included more often pleasant emotions while those of men politicians included more anger displays.

Moreover, candidates' electoral communication is affected by the interplay of their gender and party identity (Bauer, 2018). In quantitative terms, women's underrepresentation—both on electoral lists and in elected positions—tends to be stronger in conservative parties. This pattern holds true for Switzerland, with significantly higher shares of women among the elected national councilors in progressive parties (SP = 58.5; Grüne = 56.5; GLP = 70.0) compared to conservative parties (SVP = 19.4, Mitte = 31.0; FDP = 42.9; Federal office of Statistics, 2024). Existing studies indicate that the stereotypical gender identities of politicians gain particular prominence in conservative parties that champion traditional family roles (e.g., Bauer, 2018; Pradel, 2021). Conversely, politicians from left-wing parties may avoid emphasizing gender roles to maintain broader appeal with their respective electorate (Huddy & Terkildsen, 1993). Crucially, a recent analysis of Wikipedia and Google searches for German candidates showed that search engines reinforce these party-driven gender disparities (Pradel, 2021).

In sum, we contend that search engines consolidate the universe of available visual material about candidates in ways that uphold gender stereotypes.[1]

*H3*: Image searches will underrepresent women (compared to their share on lists).

---

[1] We will also explore patterns in candidates with varying party affiliations and report the analyses in section B and C of the Online Appendix.

*H4*: Image searches for women candidates will show images with (a) more smiling, (b) more positive but (c) less negative emotional displays than searches for men candidates.

*H5*: Women candidates whose image search output is more negative receive less electoral support

**Methodology**

**Data collection**

This study consists of an algorithm audit of potential biases in Google search results during Switzerland's 2023 Federal Elections. Given Switzerland's semi-direct democracy, where frequent referendums require voters to make informed decisions (Trechsel & Kriesi, 1996), search engines likely play a critical role in shaping political information access. As the dominant search platform in Switzerland, Google is likely to influence how voters learn about candidates and ballot issues (Blassnig et al., 2023).

Algorithm auditing is a research method used to systematically analyze the performance of complex decision-making systems to evaluate their functionality and impact (Mittelstadt, 2016). This method is often applied to study the performance of search engines (Robertson et al., 2018; Hu et al., 2019; Urman et al., 2022; Kuznetsova et al., 2024) as well as other algorithmic systems such as news recommenders (Bandy & Diakopoulos, 2020). To implement the audit, we used queries corresponding to the names of all candidates (N=5,883) running in the 2023 elections. The names of the candidates were extracted from the website of the Federal Statistical Office.

To conduct the audit, we developed a cloud-based infrastructure for deploying a large number of virtual agents, namely automated scripts simulating human browsing behavior (Ulloa et al., 2024c). The advantage of virtual agents is that they allow generating inputs for algorithmic systems (e.g., entering text queries or clicking links) under controlled conditions and then recording the outputs for subsequent analysis. Besides, our infrastructure allows the simultaneous deployment of multiple agents to isolate the effect of time that can affect search results and control for the randomization of search outputs, which can result in the variation in the search outputs generated for the same queries around the same time.

We programmed the agents to retrieve the top search results for Google text and image search for each search query (i.e., candidate first and last name). We used the local version of Google (i.e., google.ch) and conducted two rounds of data collection: on 1 October and 17 October 2023, namely three and one week before the elections. The timing of the data collection allowed us to assess how Google search results change over the critical phase of the electoral campaign. After the data collection, we exported the data to the university's internal server for long-term storage and analysis.

For the technical implementation, we used Google Compute Engine to deploy one storage server to preserve data during data collection and 120 virtual machines (with Zürich IPs). Each virtual machine hosted two virtual agents using the Firefox browser; each engine consisted of a script modelling the human browsing behavior (i.e., opening the search engine page and entering the search query) and then sending the resulting HTML to the storage server, while holding key parameters of the search constant (e.g., virtual machines within a particular IP range and with the same system parameters). To minimize the impact of personalization, agents were programmed to close the web browser after each query and cleaned cookies.

**Data analysis**

*Data processing*

To analyze collected data, we first built customized HTML parsers to differentiate between advertised and organic content on Google's search pages and to extract organic text and image search results. The number of organic search results is reported in Table 1. These results were then filtered to retain unique sources from the text search and images from the image search, which were then enriched for the analysis. To enrich text search results, we first matched unique source domains with the existing lists of Swiss and international media outlets. Although these lists allowed for automated identification of a large proportion of journalistic media sources, several thousand unique domains remained unknown. To address this problem, we developed a simple codebook (see Table S1 in the Online Appendix) to manually code the types of the remaining sources. After two rounds of training, a student assistant determined for each domain in the text search result whether it represented a news media source or one of nine other source types.[2]

Table 1. Overview of collected data from the two algorithm audits

|  | Wave 1 | Wave 2 |
|---|---|---|
| *Text search* | | |
| Total text results (= $n$; number of information sources) | 594'126 | 644,558 |
| Average sources per candidate (= $n/k$) | 101 | 110 |
| *Image search* | | |
| Total image search results (= $n$; number of images) | 1'805'908 | 3'635'707 |
| Average images per candidate (= $n/k$) | 308 | 619 |

*Notes*. $k$ = 5,883 queries (i.e., candidate names).

To enrich image search results, we used computer vision techniques that allow the automatic labelling of visual content items. Due to the infeasibility of manual coding of such large image corpora, we processed images using Amazon Rekognition API, a state-of-the-art computer vision tool that automatically recognizes objects (e.g., human bodies or buildings) and has extensive capacities for analyzing facial expressions. This approach allowed us to examine the affect-related representation of the candidates by detecting the emotions the candidates' images display, including: happiness, calmness, surprise, confusion, fear, sadness, anger, and disgust.

---

[2] See the codebook and the additional descriptive tables and figures in the Online Appendix for details.

*Measures*

We aggregated all data from the level of the individual text or image search result to the level of the individual candidate running for the election, resulting in separate datasets for wave 1 and 2 (each n = 5'883).

For the text searches, we used the coded sources and their rank in the search results to measure source type prominence as the rank-weighted share of each source type. Unlike raw (unweighted) shares, which treat all results equally, this measure assigns greater weight to higher-ranked results, reflecting their increased visibility and likelihood of user engagement. This adjustment is justified by empirical evidence suggesting that users disproportionately click on top-ranked links (e.g., Granka et al., 2004). Among these sources, we focus specifically on *media source prominence* as an operationalization of the visibility of news and media sources when searching for candidate-related information. Mathematically, media source prominence for a given candidate is calculated as the sum of inverse ranks for all media results normalized by the total inverse ranks across all sources:

$$Media\ source\ prominence = \frac{\sum_{i \in Media} \frac{1}{rank_i}}{\sum_{j=1}^{N} \frac{1}{rank_j}}$$

where $rank_i$ is the position of the *i*-th media result, and *N* is the total number of results. For instance, if a candidate's top three results include two news articles (ranks 1 and 3) and one government source (rank 2), media source prominence (72%) exceeds the raw media share (67%) because the top-ranked news result receives disproportionate weight. This metric isolates the competitive advantage of highly visible media sources in search rankings, accounting for both their frequency and strategic positioning.

For the image searches, we derived four measures of visual stereotyping (see also Jungblut & Haim, 2023). First, we quantified *women's representation* by calculating the average proportion of women relative to all depicted persons across images returned per candidate query. Second, using a computer vision algorithm (see above), we computed the average probability (0 to 1) that the most prominent face in each image displayed a *smile*. Third, using the same logic, we operationalized *positive affect* as the sum of probabilities for positively valenced emotional displays—happiness and calmness—and calculated its average presence across images. Fourth, we repeated this approach for *negative affect*, defined as the summed probabilities of fear, anger, sadness, and disgust expressions, yielding an average score per query

Finally, we merged the text and image search measures with publicly available election data. The primary dependent variable was candidates' log-transformed *personal vote*. From the same dataset, we extracted candidate *gender* (0 = men, 1 = women) and *party affiliation* as categorical independent variables. Additionally, we included three covariates: candidates' *list position* (normalized within cantons to account for varying ballot lengths), *incumbency* status (coded as 0 for non-incumbents and 1 for current parliamentary officeholders), and the *canton* (i.e. the geographical electoral district) in which the candidates ran for office.

*Data analysis strategy*

The data analysis consisted of two steps. First, we assessed gender and party patterns in each text and image search measure (*H1, H3, H4*). For each measure, we fitted generalised linear mixed-effects regression models with by-canton random intercepts to predict the value of each measure using candidates' gender and party affiliation as key predictors. We clustered the standard errors around the canton—that is, the regional administrative unit at which elections are organized—to account for within-region correlations due to geographical particularities. We adjusted all estimates by candidates' list position and incumbency, two known indicators of candidates' political experience and prior success. We repeated each model for both waves of data collection as an indicator of the temporal robustness of results.

Second, we then explored how candidate characteristics, text search measures, and image search measures predicted their electoral performance. Using the same generalized mixed-effects model with by-canton random intercepts, we conducted hierarchical regression analysis where we added candidate characteristics, text, and image measures as separate blocks to evaluate their explanatory power. We tested our expectations regarding the impact of media source prominence (*H2*) and visual stereotyping (*H5*) on the full models, including all variables. The full regression tables are provided in section B of the Online Appendix.

## Results

**Text search results**

In both data collection waves, two types of sources dominated candidate-related text searches (see Figures S1-S3 in the Online Appendix for a detailed breakdown). On the one hand, sources referring to administration and government websites comprised 35.1% of all sources in wave 1 and 40.0% in wave 2, mainly including pages belonging to local or federal websites, or official election platforms. On the other hand, and as expected, media sources constituted the other main category of source results. Approximately one-third of results for candidate queries were constituted by legacy and online news media websites (33.7% in wave 1 and 29.7% in wave 2). Official party websites were the third most common source (12.0% in wave 1 and 14.0% in wave 2). Differences between data collection waves were minimal, suggesting relatively stable online voter information.

We next investigated what candidate characteristics explained the rank-weighted visibility of media-related search results, expecting higher media source prominence for men candidates (*H1a*) and for candidates running for larger parties (*H1b*). Panel A of Figure 1 shows the predicted media source prominence for women and men candidates. For both waves, results indicated significantly higher media source prominence for men than women candidates (wave 1: *b* = -3.94, 95%CI = -4.83 – -3.04, *p* < .001; wave 2: *b* = -2.72, 95%CI = -3.68 – -1.76, *p* <.001). In line with our expectation (*H1a*), Google systematically included more (and higher ranked) sources to news media for men candidates, granting them more media visibility in text searches.

Party differences were less clear. Panel B of Figure 1 shows the predicted media source prominence for all major parties (ranked in terms of their total electoral shares from left to right. Against our expectation (*H1b*), candidates from the largest party, the right-wing conservative Swiss People's Party (SVP), had significantly lower media source prominence compared to candidates from all other parties. Yet the Swiss People's Party appears to be an exception. Post hoc pairwise comparisons with Benjamini-Hochberg correction for multiple comparisons revealed no significant difference between the second to fourth largest parties (Social Democratic Party[SP], Liberal Party[FDP], Center Party [Mitte]). However, all contrasts were significant between each of these parties and the remaining, substantially smaller parties (Green Party [GRÜNE], Green Liberal Party [GLP], and pooled smaller parties [OTHER]). Overall, these results lend only partial support to the idea of a positive association between party size and increased media source prominence on search engines.

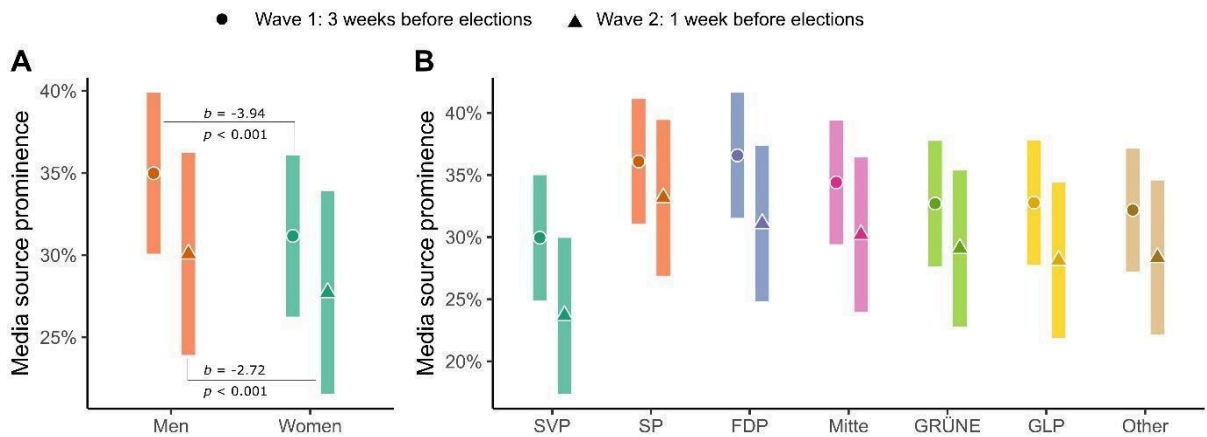

**Figure 1**. Share (Panel A) and ranking (B) of sources in Google text searches for women and men candidate queries in the run-up to the Swiss Federal Elections. Note: Estimates present aggregated means across candidate queries and virtual agents along with their 95% confidence intervals. Categories with low frequencies were omitted due to random measurement errors possibly being higher than their actual occurrence.

**Image search results**

*Quantitative representation*

Turning to the measures derived from Google image searches, we first investigated the quantitative dimension of visual stereotyping by predicting the share of women shown in images of search results. Panel A of Figure 2 illustrates the distribution of the share of women, showing that, on average, they present 38.7% of depicted persons in the Google image search (SD = 13.5) in wave 1 and 39.1% (SD = 13.2) in wave 2. Women's visual representation on Google is in both waves significantly lower than their representation of official electoral lists (42.5%), although only by a few percentage points (wave 1: $t(13)$ = -10.2, $p < .001$; wave 2: $t(13)$ = -9.2, $p < .001$). Though providing support for our expectation of women's visual underrepresentation (*H3*), the extent of gender bias remained rather small.

To further probe the extent of women's underrepresentation, we took a closer look at the gender composition of Google images for queries of women and men candidates by predicting the raw gender difference in the number of men from the number of depicted women for each query (see Panel B of Figure 2). Interestingly, the gender difference was statistically indistinguishable from zero for queries of women candidates in both waves (wave 1: $b$ = -0.12, 95%CI = -0.28 – 0.06, $p$ = .19; wave 2: $b$ = -0.11, 95%CI = -0.29 – 0.07, $p$ = .23). In other words, when voters specifically searched for women on Google, they encountered as many men as women in the image results. Crucially, the same pattern did not apply to queries of men candidates whose queries resulted, on average, in images that represented significantly more men than women (wave 1: $b$ = -0.98, 95%CI = -1.15 – -0.81, $p$ < .001; wave 2: $b$ = -0.94, 95%CI = -1.12 – -0.76, $p$ < .001). These results suggest that men politicians were by default visible on search engines, while women only reached similar levels of visibility for specific user queries.

Finally, we found a (small) positive correlation between the share of women on official lists and their representation on Google images, $r(558)$ = .12, $p$ = .002 (see Panel C in Figure 2). This association was unsurprising, as much of the visual campaign materials are organized at the level of lists; therefore, women who ran together with more women in the electoral campaign were also more likely to appear together online. Yet this finding is not trivial, as it showcases the impact of parties' offline campaigning decisions (e.g., having quotas on lists or running all-women lists) on candidates' online presence.

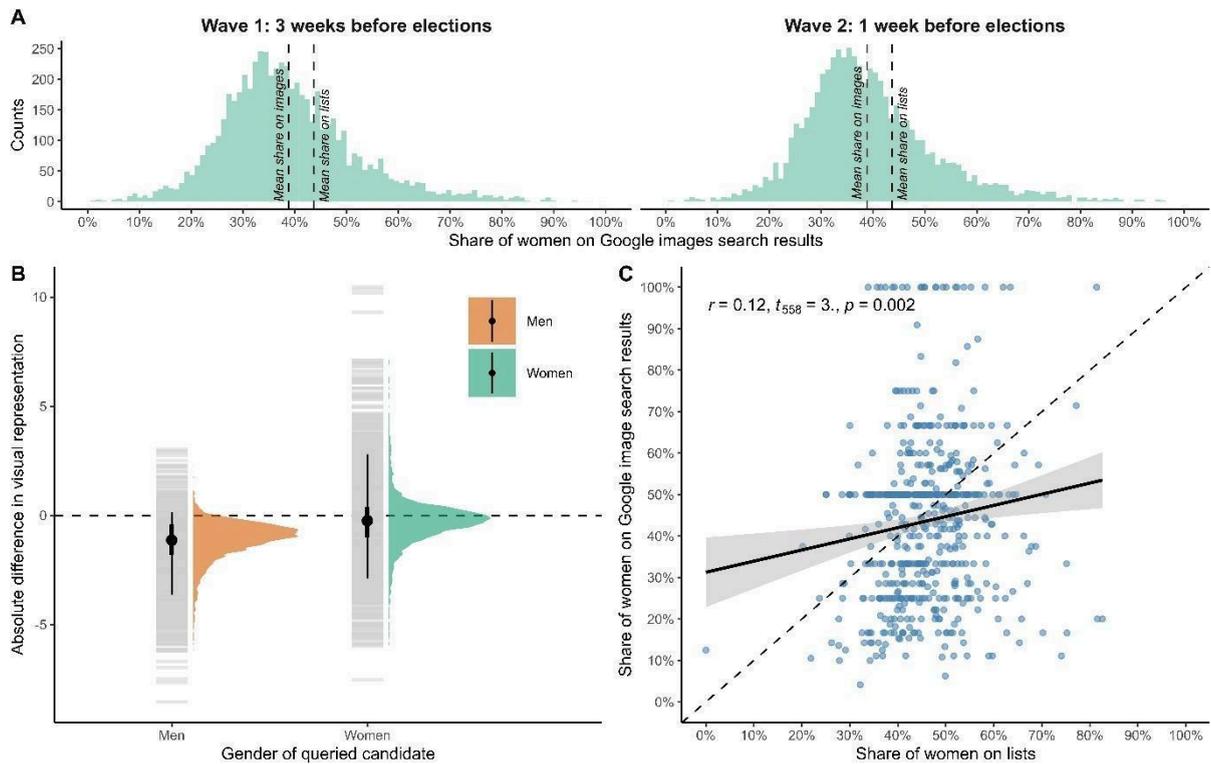

**Figure 2.** Overview of women's representation on Google image searches for candidate queries in the run-up to the Swiss Federal Elections. Panel A depicts the distribution of women's representation across both waves of data collection. Panel B shows the absolute difference in the number of men and women shown in the image results of queries for men and women candidates. Panel C shows the bivariate correlation between women's representation on electoral lists and their average representation on image searches.

*Emotional displays*

We first examined the frequency of images showing candidates smiling (see Panel A of Figure 3). In line with our expectation (*H4a*), image search results revealed that women candidates exhibited a higher prevalence of smiling images, with an average of 65.3% (SD = 19.4) compared to 60.9% for men candidates (SD = 18.7; see Panel A of Figure 6). However, the average gender difference in smiling is rather small, with 5.4% percentage points (95%CI = 4.49 – 6.36, $p < .001$). Across queries, we found that emotional expressions conveying positive affect ($M = 55.9\%$, $SD = 15.5$) are much more common than those conveying negative affect ($M = 14.5\%$, $SD = 6.4$). Moreover, the prevalence of affective displays varies across candidate gender consistent with our expectations, with queries for women candidates resulting in images with significantly higher average positive affect (wave 1: $b = 1.82$, 95%CI = 1.02 – 2.62, $p < .001$; wave 2: $b = 2.22$, 95%CI = 1.42 – 3.00, $p < .001$; see *H4a*) and lower negative affect (wave 1: $b = -1.05$, 95%CI = -1.37 – -0.72, $p < .001$; wave 2: $b = -1.08$, 95%CI = -1.37 – -0.79, $p < .001$; see *H4b*).

Furthermore, we explored how party affiliation affects candidates' affective displays. Candidates from more left-leaning parties (GRÜNE, SP) tended to have a higher (lower) prevalence of negative (positive) affect in image search results than those from conservative parties (SVP, FDP) across both waves. As part of this exploration, we also repeated the models with an additional gender-by-party interaction term. The results are depicted in Panel B of Figure 3, suggesting that Google images tended to depict women candidates from conservative parties with particularly more positive and less negative affect compared to their male counterparts and women from progressive parties. For instance, image searches for women from the GRÜNE and SP yield identical prevalence rates of negative affect as those for men candidates running for both conservative and progressive parties.

To evaluate the robustness of gendered patterns across regional electoral contexts, we aggregated the affective displays as standardized mean differences between women and men candidates for each canton. Positive values indicate higher affect shares in images of women, while negative values indicate lower shares. Panel C of Figure 3 presents the results with three key insights. First, most cantons exhibited no gender differences in affect-related representation, as indicated by the grey bars. Second, the skewness in gender affect displays was not uniformly aligned; in only three cantons (Bern, Lucerne, Grisons) did Google image search results for women candidates show both more positive *and* less negative affect. Third, there were no counterexamples to these patterns; specifically, no instances where images of men display *more* positive or *less* negative affect compared to women.

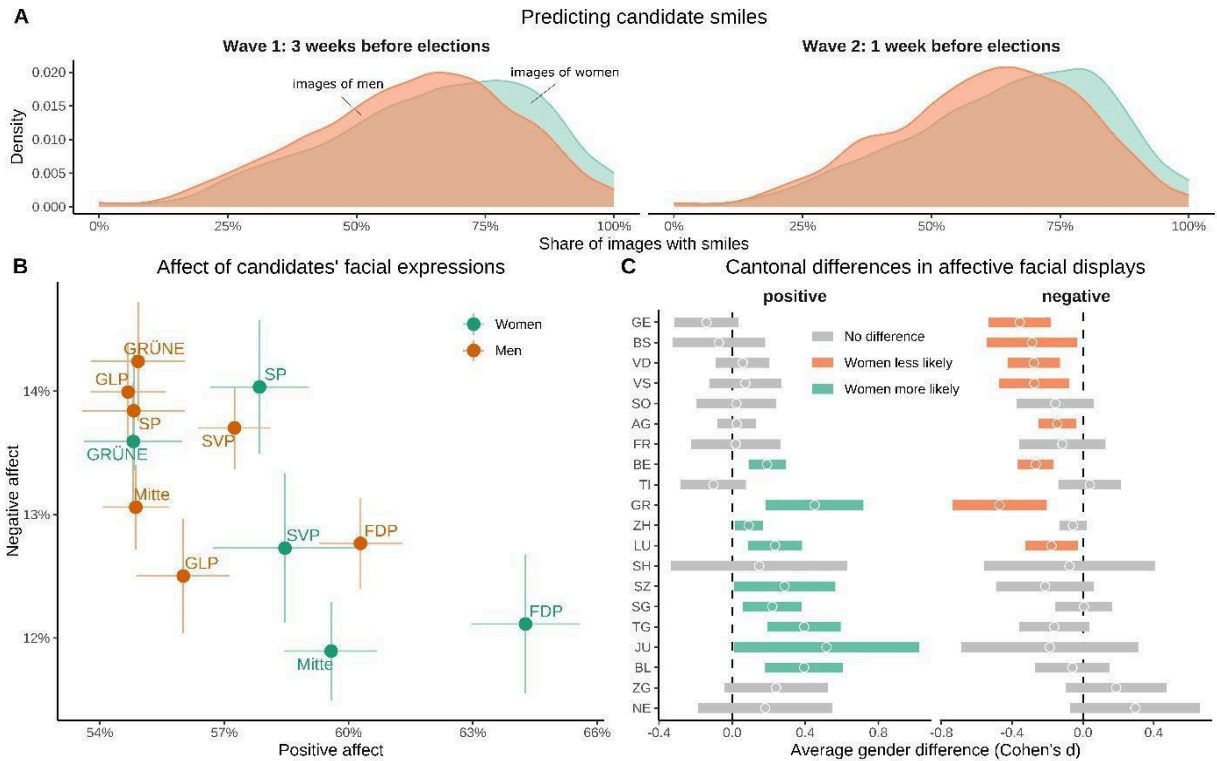

**Figure 3**. Gendered patterns in emotional displays in Google image searches. Panel A shows the distribution of images with smiling faces for queries of women and men candidates. The bottom half shows results for emotions aggregated into positive and negative affect categories by party (Panel B) and canton (Panel C). Note: Data only shows wave 1 for bottom panels. Panel B shows average marginal effects along with 95% confidence intervals. Panel C shows standardised mean differences (Cohen's *d* with Hedges' correction), split by canton. Half cantons (OW/NW) were excluded due to imprecise measurements resulting from the low number of running candidates.

*Search engines and electoral results*

A last set of analyses explored how the measures derived from Google text and image searches predicted electoral outcomes. The full models for waves 1 and 2 explained 42.9% and 41.6% of the variance in candidates' personal votes, respectively, as indicated by the conditional $R^2$ values (see Table 2). Approximately half of this explained variance was attributed to random effects, capturing the clustering of observations by canton due to differing canton sizes and other unobserved variables specific to each local race. The remaining variance was explained by three blocks of fixed effects: candidate characteristics, text search measures, and image search measures, with a total marginal $R^2$ of 22.6% for wave 1 and 22.0% for wave 2.

Traditional *candidate characteristics* accounted for 15.3% of the variation in personal votes, with candidate gender, party affiliation, incumbency status, and list position significantly predicting personal votes.

Likelihood ratio tests indicated that adding text search variables significantly enhanced model fit in both waves, $\chi^2(3) = 394.1$, $p < .001$ and $\chi^2(3) = 288.4$, $p < .001$, accounting for an additional 4.7% and 3.4% of the variance in candidates' votes, respectively. We hypothesized that candidates with higher media source prominence in Google text searches would perform better electorally (*H2*). In wave 1, media source prominence had a significant positive effect on personal votes ($b = 0.27$, 95%CI = 0.05 – 0.49, $p < .001$). Notably, a one-unit increase in media source prominence corresponded to an average 31% increase in personal votes. Similarly, in wave 2, media prominence maintained a significant positive impact ($b = 1.34$, 95%CI = 1.001 – 1.58, $p < .001$), corroborating our hypothesis. This suggests that even small differences in media prominence, such as 3–4 percentage points between women and men candidates, may have substantial electoral implications. To further investigate gendered effects, we included an interaction term in the model. The significant interaction effect in wave 1 ($b = 1.37$, 95%CI = 0.46 – 2.28, $p = .003$) is depicted in panel B of Figure 4, illustrating that women candidates particularly benefit from having more prominent media sources in Google text searches.

**Table 2.** Hierarchical mixed-effects regression models predicting log-transformed personal votes

|  | Wave 1 | | | Wave 2 | | |
| --- | --- | --- | --- | --- | --- | --- |
|  | *b* | *se* | *p* | *b* | *se* | *p* |
| *Controls* | | | | | | |
| (Intercept) | 7.49 | 0.15 | <.001 | 7.49 | 0.15 | <.001 |
| Woman | 0.19 | 0.03 | <.001 | 0.19 | 0.03 | <.001 |
| Incumbent | 2.90 | 0.10 | <.001 | 2.90 | 0.10 | <.001 |
| Party [FDP] | -0.33 | 0.07 | <.001 | -0.33 | 0.07 | <.001 |
| Party [GLP] | -0.70 | 0.07 | <.001 | -0.70 | 0.07 | <.001 |
| Party [GRÜNE] | | -0.22 | 0.07 | .002 | -0.22 | 0.07 | .002 |
| Party [Mitte] | | -0.65 | 0.06 | <.001 | -0.65 | 0.06 | <.001 |
| Party [SP] | | -0.13 | 0.07 | .054 | -0.13 | 0.07 | .055 |
| Party [Other] | | -0.78 | 0.06 | <.001 | -0.78 | 0.06 | <.001 |
| List position | | -0.14 | 0.06 | .019 | -0.13 | 0.06 | .020 |
| Marg. $R^2$ | | .153 | | | .153 | | |
| *Text search: Source prominence (+ Controls)* | | | | | | | |
| News media | | 0.27 | 0.11 | <.001 | 0.37 | 0.11 | <.001 |
| Social media | | 1.76 | 0.26 | .015 | 1.34 | 0.18 | .001 |
| Government and administration | | -1.41 | 0.11 | <.001 | -0.84 | 0.10 | <.001 |
| Δ Marg. $R^2$ | | .047 | | | .034 | | |
| | | $\chi(3) = 394.1, p < .001$ | | | $\chi(3) = 288.4, p < .001$ | | |
| *Image search: Visual stereotyping (+ Controls + text search)* | | | | | | | |
| Positive affect | | 0.01 | 0.00 | <.001 | 0.02 | 0.00 | <0.001 |
| Negative affect | | 0.01 | 0.00 | <.001 | 0.02 | 0.00 | <0.001 |
| Share of smiles | | 0.01 | 0.00 | <.001 | 0.01 | 0.00 | <0.001 |
| Share women | | -0.02 | 0.13 | .873 | -0.12 | 0.14 | .386 |
| Δ Marg. $R^2$ | | .025 | | | .031 | | |
| | | $\chi(4) = 259.4, p < .001$ | | | $\chi(4) = 310.9, p < .001$ | | |
| *Interactions (+ Controls + text search + image search)* | | | | | | | |
| Woman × Negative affect | | -0.01 | 0.00 | .248 | -0.00 | 0.01 | .484 |
| Woman × Media source prominence | | 1.37 | 0.46 | .003 | 0.10 | 0.30 | .723 |
| Δ Marg. $R^2$ | | .000 | | | .000 | | |
| | | $\chi(2) = 10.0, p = .006$ | | | $\chi(2) = 0.632, p = .729$ | | |
| **Random Effects** *(complete model)* | | | | | | | |
| Total Marg. $R^2$ / cond. $R^2$ | | .226 / .429 | | | .220 / .416 | | |
| ICC / $\sigma^2$ / $\tau_{00}$ | | .26 / 1.35 / $0.48_{canton}$ | | | .25 / 1.36 / $0.46_{canton}$ | | |
| Observations (candidates) / N | | 5952 / $22_{canton}$ | | | 5948 / $22_{canton}$ | | |

*Notes.* Model fits are tested with Likelihood ratio tests.

*Image search-related* variables further improved fit in both waves, $\chi^2(4) = 259.4, p < .001$ and $\chi(4) = 310.9, p < .001$, accounting for increases of 2.5% and 3.1% in explained variance, respectively. Interestingly, both positive and negative affective displays positively affected electoral success, suggesting candidates have some flexibility in their emotional expressions in visual campaign materials. We posited that expressions of negative emotions might contravene gender norms and lead to poorer electoral performance for women (*H5*). However, this hypothesis was not supported, as the data revealed no significant interaction term in either wave. Panel A of Figure 4 illustrates this interaction, showing that voters respond similarly to negative affect in image search queries for both men and women candidates.

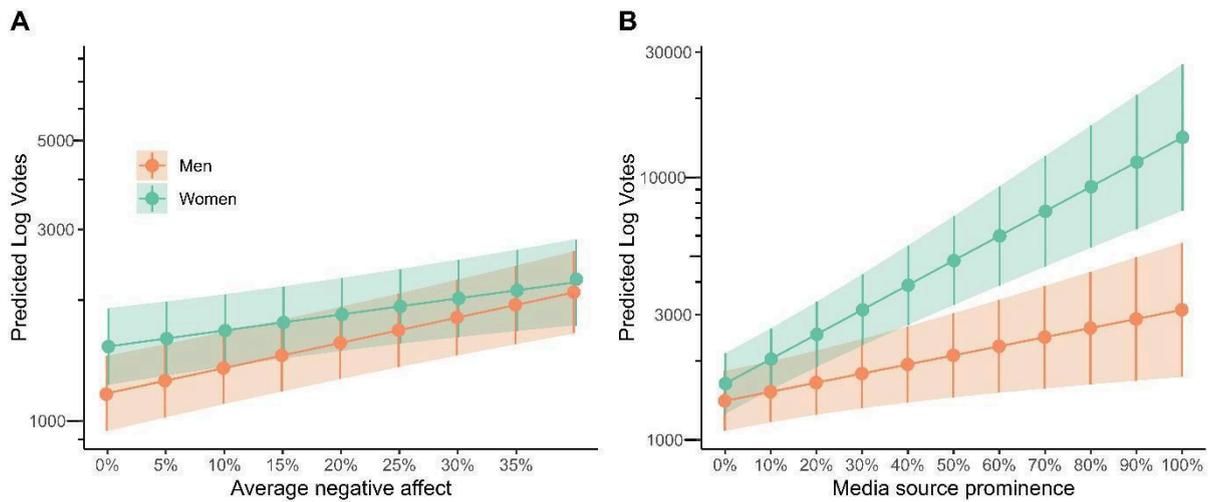

**Figure 4**. Interaction effects of gender of queried candidates and the average negative affect in image searches (Panel A) and media source prominence (Panel B) on candidates' predicted personal votes. Data show the predicted marginal effects along with 95% Confidence Intervals for wave 1 of data collection.

**Discussion and Conclusion**

In this article, we examined how Google selects and ranks information about candidates for the 2023 Federal Elections in Switzerland. By conducting an algorithm audit of text and image search results related to all the candidates participating in the elections, we aimed to understand whether Google's outputs about candidates differ according to their party and gender and whether certain candidate groups can be particularly prone to skewed representation by Google's algorithms. We then assessed candidate-level search engine output collected during the campaigning period to predict candidates' electoral performance.

The audit of Google's *text search* results revealed three key findings. First, similar to earlier studies of search engine prioritization of election-related information (Haim et al., 2018; Urman et al., 2022; Evans et al., 2023), we observed the strong prevalence of authoritative sources, namely government websites and news media. Media organizations' investment in search engine optimization techniques seems to be paying dividends (Schwab et al., 2023). At the same time, within these authoritative sources, we observed unequal treatment of politicians based on their gender. Specifically, men candidates received more and higher ranked links to news media reporting than women candidates, suggesting a bias that may reduce women's media visibility in elections. This finding expands research on women's underrepresentation in traditional political media coverage (Van der Pas & Aaldering, 2022) to include algorithmically driven search engines. Our results suggest that women miss out on a key campaigning advantage due their lower media source prominence in two ways. First, we found that news media prominence—which is lower for queries of women politicians—positively impacted candidates' electoral performance. Second, we also found that this benefit was especially pronounced for women candidates.

Surprisingly, we did not observe the expected relationship between the electoral weight of a party and its media prominence in search results, which counters the "rich get (or remain) richer" logic posited by the normalization hypothesis of political online communication (e.g., Puschmann, 2019). The results suggest that search engines may operate according to their own logic that is more than a simple amplification of traditional offline power dynamics when it comes to explaining parties' representation (Pradel, 2021). At the same, it is important to acknowledge that an absence of partisan differences in search engine output may be due to particularities of the Swiss political landscape. Specifically, the largest political party, SVP, often attracts more support in rural cantons covered by smaller regional media. Some of these media may not necessarily have digital editions or they may lack resources for advanced search engine optimization techniques that play to search engines' prioritization algorithms (see Schwab et al., 2022; Haim et al., 2018). Future research can benefit from tracing the link between media source prominence and party size across countries with different political landscapes, which is essential for understanding the possible impacts of search engines on electability of candidates in the context of elections.

The results for *image search* results tend to reaffirm the pervasiveness of gender stereotypes in political communication. In quantitative terms, we observed a slight underrepresentation of women in image searches, which is less pronounced compared with might have been expected based on earlier research (e.g., Otterbacher et al., 2017; Guilbeault et al., 2024). This observation suggests that the degree of quantitative gender bias (at least in terms of quantitative representation) in search engines might be closely connected to the political situation of women in the context under study (Rohrbach et al., 2024). As women are becoming part of the political fabric, gender representation on search engines might shift from a difference to a similarity model (Rohrbach et al., 2023).

    In more qualitative terms, we found that the images associated with women candidates, especially from conservative parties, more often focused on positive and warm emotions. On the one hand, these patterns align with gender stereotypes, suggesting the continued influence of gender norms on candidates' self-presentation. In contrast, candidates from progressive parties may challenge traditional gender norms, resulting in a broader range of emotional portrayals that do not conform to stereotypical expectations. One interpretation is that search engines can be seen as reinforcing women's double-bind in politics by highlighting traits that are seemingly at odds with political leadership (see Aaldering et al., 2018; Rohrbach et al., 2023).  On the other hand, the emphasis of the communality stereotype for representing women candidates in digital information landscapes may also serve a strategic purpose (see Haim & Jungblut, 2023), conveying warmth rather than solely competence. This is line with recent research highlighting that women (but not men) can effectively leverage a "feminine advantage" by focusing on communal aspects in their communication (Bauer, 2020; Rohrbach, 2024). In this case, Google's image search performance may facilitate representation strategies that can benefit women politicians, even if any such advantage may further cement prescriptive gender norms.

    Taken together, these observations stress the importance of search engines in political communication processes and demonstrate the ambiguous role played by different forms of information searches. While the idea of Google being a crucial gatekeeper of political

information is not new, our study empirically shows that various measures derived from how text and image search results represent information about politicians can be linked to electoral performance. Together, measures from search engines explained between 6–8% of the variance in voting during the Swiss 2023 Federal Elections, which is substantial given that elections are often decided over differences of just a few percentage points. This observation is particularly concerning if we consider that our findings also highlight that Google is far from being an equal playing field for candidates, as evidenced by the differences in how text and image searches treat candidates from different gender groups and from different parties. Whereas traditional journalistic gatekeeping can be studied by studying journalists and their organizational contexts, the "black box" character of algorithmic systems makes it difficult to understand what influences drive search engines as the new "last" or "terminal" gate (Bro & Wallberg, p. 452; Wallace, 2018). This study extends the line of work demonstrating the usefulness of algorithm audits as a method to reconstruct performance in digital gatekeeping processes (Bandy & Diakopoulos, 2020), making them an integral part of the methodological toolkit of digital journalism research.

  Our findings raise several urgent questions for researchers and regulators. There is a need to explore how to distinguish algorithmic bias from social bias. While our study shows that Google's search algorithms treat women and women candidates differently, it's unclear whether these algorithms merely amplify existing social biases or create distortions in social reality (Rohrbach et al., 2024). Establishing a robust baseline for search engine curation each politician—by disentangling strategic self-presentation, representation through media and search engines, and political context—is essential but challenging (for an example, see Haim & Jungblut, 2023). Another important question pertains to the normative expectations of search engines' performance concerning political information. Although research indicate that users generally accept algorithmic curation of information (Thurman et al., 2019), it remains an open question to what extent audiences support forms of "debiasing" curation designed to minimize discrepancies in candidate representation (Wang et al., 2022).

Addressing these questions is essential for shaping a long-term vision of search engines' roles in democratic decision-making. This aligns with Puschmann's (2019) dual characterization of search engines in digital journalism research. On one hand, it underscores the necessity of empirical studies that examine how search engines are *influencing politics* through their (mal)performance across various countries and themes. On the other hand, such evidence can stimulate public discourse about search engines *having politics* and inform potential new regulations involving a diverse array of stakeholders, including administration, media, political parties, civil society, industry, and academia.

# Online Appendix

Campaigning through the lens of Google: A large-scale algorithm audit of Google searches in the run-up to the Swiss Federal Elections 2023

## Content

# A: Codebook

We coded the news sources in the text search output at the level of the URLs (z.B. www.Google.ch/…). Only in cases the URL does not allow for coding, coders clicked on the link to get more information about the source (e.g. in the "about" tab)

**Table S1**. Codebook of sources in text searches

*Media sources*
1. **News, media, election coverage**
   (Example: *srf.ch, 20min.ch, infosperber.ch, nzz.ch, rts.ch, tio.ch, arcinfo.ch, medienwoche.ch, persoenlich.ch etc.*)
2. **Social media**
   (Example: *linkedin.com, twitter.com, x.com, facebook.com, instagram.com, YouTube etc.*)

*Political sources*
3. **Party website**
   (Example: *sp-ps.ch, fdp.ch, plr.ch, plrt.ch, svp.ch, it.udc.ch, svp.ch etc.*)
4. **Personal candidate websites**
   (Beispiele: *reginesauter.ch, gerhard-andrey.ch, aeschi.com, thierry-burkhard.ch etc.*)
5. **Government websites** (all levels of office)
   (Example: *parlament.ch, ch.ch, be.ch, bern.ch, bfs.admin.ch, bag.admin.ch, admin.ch etc.*)
6. **Political think tanks and movements**
   (Example: *mass-voll.ch, foraus.ch, amnesty.ch, algorithmwatch.org, avenir-suisse.ch etc.*)
   **Note**: In case of doubt, visit website and read "about us" page.
7. **Vote recommenders**
   (Example: *parteienkompass.ch, smartvote.ch, easyvote.ch etc.*)

*Other sources*
8. **Commercial websites from business, industry & companies**
   (Example: *kellerhals-carrard.ch, nestle.ch, raiffeisen.ch, economiesuisse.ch etc.*)

   **Note:** This refers mainly to URLs that pertain to the economic networks/mandates and employers (SMEs, law firms, practices, etc.) of candidates
9. **Encyclopedia/fora**
   (Example: *wikipedia.org, quora.com etc.*)
10. **Miscellaneous**
    **Note:** In this case, please provide a few words about the impression of the source in the comments field

# B: Full results of reported analyses

## H1

**Table S2.** Predicting media source prominence

|  | Wave 1 | | | Wave 2 | | |
|---|---|---|---|---|---|---|
| *Predictors* | b | CI | p | b | CI | p |
| (Intercept) | 36.57 | 31.49 – 41.65 | **<0.001** | 27.24 | 20.93 – 33.55 | **<0.001** |
| Woman | -3.94 | -4.83 – -3.04 | **<0.001** | -2.72 | -3.68 – -1.76 | **<0.001** |
| Incumbent | 19.05 | 16.34 – 21.76 | **<0.001** | 19.93 | 17.04 – 22.81 | **<0.001** |
| Party [FDP] | 4.52 | 2.62 – 6.41 | **<0.001** | 3.61 | 1.59 – 5.62 | **<0.001** |
| Party [GLP] | 5.17 | 3.37 – 6.96 | **<0.001** | 5.32 | 3.41 – 7.23 | **<0.001** |
| Party [GRÜNE] | 3.51 | 1.61 – 5.40 | **<0.001** | 5.12 | 3.10 – 7.14 | **<0.001** |
| Party [Mitte] | 4.88 | 3.19 – 6.58 | **<0.001** | 5.31 | 3.50 – 7.11 | **<0.001** |
| Party [Other] | 7.18 | 5.53 – 8.83 | **<0.001** | 7.03 | 5.27 – 8.79 | **<0.001** |
| Party [SP] | 6.58 | 4.72 – 8.44 | **<0.001** | 7.98 | 5.99 – 9.96 | **<0.001** |
| List position | 1.50 | -0.08 – 3.08 | 0.063 | 1.69 | 0.01 – 3.37 | **0.049** |
| **Random Effects** | | | | | | |
| $\sigma^2$ | 290.26 | | | 328.42 | | |
| $\tau_{00}$ | 128.45 $_{canton}$ | | | 205.22 $_{canton}$ | | |
| ICC | 0.31 | | | 0.38 | | |
| N | 22 $_{canton}$ | | | 22 $_{canton}$ | | |
| Observations | 5996 | | | 5989 | | |
| Marginal $R^2$ / Conditional $R^2$ | 0.038 / 0.333 | | | 0.030 / 0.403 | | |

# H3

Table S3. Predicting the share of women's representation in Google image searches

| Predictors | Wave 1 | | | | | | Wave 2 | | | | | |
|---|---|---|---|---|---|---|---|---|---|---|---|---|
| | Estimates | std. Error | p | Estimates | std. Error | p | Estimates | std. Error | p | Estimates | std. Error | p |
| (Intercept) | 38.78 | 0.36 | <0.001 | 28.81 | 0.57 | <0.001 | 39.05 | 0.37 | <0.001 | 29.36 | 0.59 | <0.001 |
| Woman | | | | 13.62 | 0.31 | <0.001 | | | | 12.81 | 0.31 | <0.001 |
| Incumbent | | | | -3.57 | 1.18 | **0.003** | | | | -3.65 | 1.18 | **0.002** |
| Party [FDP] | | | | 4.06 | 0.64 | <0.001 | | | | 4.00 | 0.64 | <0.001 |
| Party [GLP] | | | | 4.76 | 0.60 | <0.001 | | | | 4.27 | 0.60 | <0.001 |
| Party [GRÜNE] | | | | 5.55 | 0.64 | <0.001 | | | | 5.46 | 0.64 | <0.001 |
| Party [Mitte] | | | | 5.02 | 0.57 | <0.001 | | | | 5.00 | 0.57 | <0.001 |
| Party [Other] | | | | 4.22 | 0.56 | <0.001 | | | | 3.69 | 0.55 | <0.001 |
| Party [SP] | | | | 6.63 | 0.63 | <0.001 | | | | 6.03 | 0.63 | <0.001 |
| List position | | | | 0.18 | 0.53 | 0.729 | | | | 0.85 | 0.53 | 0.107 |

| **Random Effects** | | | | |
| --- | --- | --- | --- | --- |
| $\sigma^2$ | 181.22 | 127.09 | 174.61 | 126.47 |
| $\tau_{00}$ | 1.63 $_{canton}$ | 1.21 $_{canton}$ | 1.70 $_{canton}$ | 1.63 $_{canton}$ |
| ICC | 0.01 | 0.01 | 0.01 | 0.01 |
| N | 22 $_{canton}$ | 22 $_{canton}$ | 22 $_{canton}$ | 22 $_{canton}$ |
| Observations | 5982 | 5982 | 5986 | 5986 |
| Marginal $R^2$ / Conditional $R^2$ | 0.000 / 0.009 | 0.298 / 0.305 | 0.000 / 0.010 | 0.274 / 0.284 |

# H4

**Table S4.** Predicting average positive affect in Google image searches

| Predictors | affect positive | | | affect positive | | |
|---|---|---|---|---|---|---|
| | Estimates | CI | p | Estimates | CI | p |
| (Intercept) | 56.21 | 54.16 – 58.25 | **<0.001** | 57.35 | 55.28 – 59.43 | **<0.001** |
| Woman | 1.82 | 1.02 – 2.62 | **<0.001** | 2.22 | 1.43 – 3.00 | **<0.001** |
| Incumbent | 7.57 | 5.16 – 9.98 | **<0.001** | 9.16 | 6.79 – 11.54 | **<0.001** |
| Party [FDP] | 3.53 | 1.85 – 5.22 | **<0.001** | 4.15 | 2.49 – 5.81 | **<0.001** |
| Party [GLP] | -1.78 | -3.38 – -0.19 | **0.028** | -2.04 | -3.61 – -0.47 | **0.011** |
| Party [GRÜNE] | -3.23 | -4.92 – -1.55 | **<0.001** | -3.64 | -5.31 – -1.98 | **<0.001** |
| Party [Mitte] | -1.05 | -2.55 – 0.46 | 0.174 | -0.77 | -2.26 – 0.71 | 0.305 |
| Party [Other] | -3.02 | -4.49 – -1.55 | **<0.001** | -3.21 | -4.65 – -1.76 | **<0.001** |
| Party [SP] | -0.96 | -2.62 – 0.69 | 0.254 | -1.77 | -3.40 – -0.14 | **0.034** |
| List position | 2.06 | 0.65 – 3.46 | **0.004** | 1.57 | 0.19 – 2.96 | **0.025** |
| **Random Effects** | | | | | | |
| $\sigma^2$ | 230.22 | | | 222.95 | | |
| $\tau_{00}$ | 12.21 $_{canton}$ | | | 13.10 $_{canton}$ | | |
| ICC | 0.05 | | | 0.06 | | |
| N | 22 $_{canton}$ | | | 22 $_{canton}$ | | |
| Observations | 5996 | | | 5998 | | |
| Marginal $R^2$ / Conditional $R^2$ | 0.027 / 0.076 | | | 0.036 / 0.089 | | |

**Table S5.** Predicting average negative affect in Google image searches

|  | Wave 1 | | | Wave 2 | | |
|---|---|---|---|---|---|---|
| Predictors | Estimates | CI | p | Estimates | CI | p |
| (Intercept) | 14.32 | 13.11 – 15.53 | **<0.001** | 12.84 | 11.78 – 13.89 | **<0.001** |
| Woman | -1.05 | -1.37 – -0.72 | **<0.001** | -1.08 | -1.37 – -0.79 | **<0.001** |
| Incumbent | 3.01 | 2.03 – 4.00 | **<0.001** | 2.69 | 1.82 – 3.56 | **<0.001** |
| Party [FDP] | 0.63 | -0.06 – 1.32 | 0.074 | -1.18 | -1.78 – -0.57 | **<0.001** |
| Party [GLP] | 1.26 | 0.60 – 1.91 | **<0.001** | -0.23 | -0.81 – 0.34 | 0.432 |
| Party [GRÜNE] | 1.53 | 0.84 – 2.22 | **<0.001** | 0.33 | -0.27 – 0.94 | 0.281 |
| Party [Mitte] | 0.77 | 0.15 – 1.38 | **0.015** | -0.95 | -1.50 – -0.41 | **0.001** |
| Party [Other] | 1.16 | 0.56 – 1.76 | **<0.001** | -0.12 | -0.65 – 0.41 | 0.659 |
| Party [SP] | 2.06 | 1.38 – 2.73 | **<0.001** | 0.18 | -0.41 – 0.78 | 0.546 |
| List position | 0.39 | -0.18 – 0.97 | 0.181 | 0.16 | -0.34 – 0.67 | 0.524 |
| **Random Effects** | | | | | | |
| $\sigma^2$ | 38.47 | | | 29.88 | | |
| $\tau_{00}$ | 6.13 $_{canton}$ | | | 4.64 $_{canton}$ | | |
| ICC | 0.14 | | | 0.13 | | |
| N | 22 $_{canton}$ | | | 22 $_{canton}$ | | |
| Observations | 5996 | | | 5998 | | |
| Marginal $R^2$ / Conditional $R^2$ | 0.016 / 0.151 | | | 0.020 / 0.152 | | |

# Predicting smiles

**Table S6.** Predicting average share of smiling faces in Google image searches

|  | first person smile*100 | | | first person smile*100 | | |
|---|---|---|---|---|---|---|
| Predictors | Estimates | CI | p | Estimates | CI | p |
| (Intercept) | 62.81 | 57.88 – 67.73 | **<0.001** | 61.54 | 57.04 – 66.03 | **<0.001** |
| Woman | 5.43 | 4.49 – 6.36 | **<0.001** | 5.68 | 4.77 – 6.58 | **<0.001** |
| Incumbent | -3.56 | -6.38 – -0.74 | **0.013** | -2.36 | -5.09 – 0.37 | 0.090 |
| Party [FDP] | -0.34 | -2.31 – 1.63 | 0.735 | 2.45 | 0.54 – 4.35 | **0.012** |
| Party [GLP] | -5.04 | -6.91 – -3.17 | **<0.001** | -4.83 | -6.64 – -3.02 | **<0.001** |
| Party [GRÜNE] | -6.53 | -8.50 – -4.55 | **<0.001** | -5.46 | -7.37 – -3.55 | **<0.001** |
| Party [Mitte] | -2.66 | -4.42 – -0.89 | **0.003** | -1.34 | -3.05 – 0.36 | 0.123 |
| Party [Other] | -6.77 | -8.49 – -5.05 | **<0.001** | -5.31 | -6.98 – -3.65 | **<0.001** |
| Party [SP] | -7.84 | -9.78 – -5.90 | **<0.001** | -7.44 | -9.32 – -5.57 | **<0.001** |
| List position | 0.80 | -0.84 – 2.45 | 0.340 | 0.67 | -0.92 – 2.26 | 0.410 |
| **Random Effects** | | | | | | |
| $\sigma^2$ | 312.30 | | | 292.78 | | |
| $\tau_{00}$ | 118.71 $_{canton}$ | | | 96.99 $_{canton}$ | | |
| ICC | 0.28 | | | 0.25 | | |
| N | 22 $_{canton}$ | | | 22 $_{canton}$ | | |
| Observations | 5952 | | | 5948 | | |
| Marginal $R^2$ / Conditional $R^2$ | 0.029 / 0.296 | | | 0.037 / 0.276 | | |

# C: Additional tables and figures

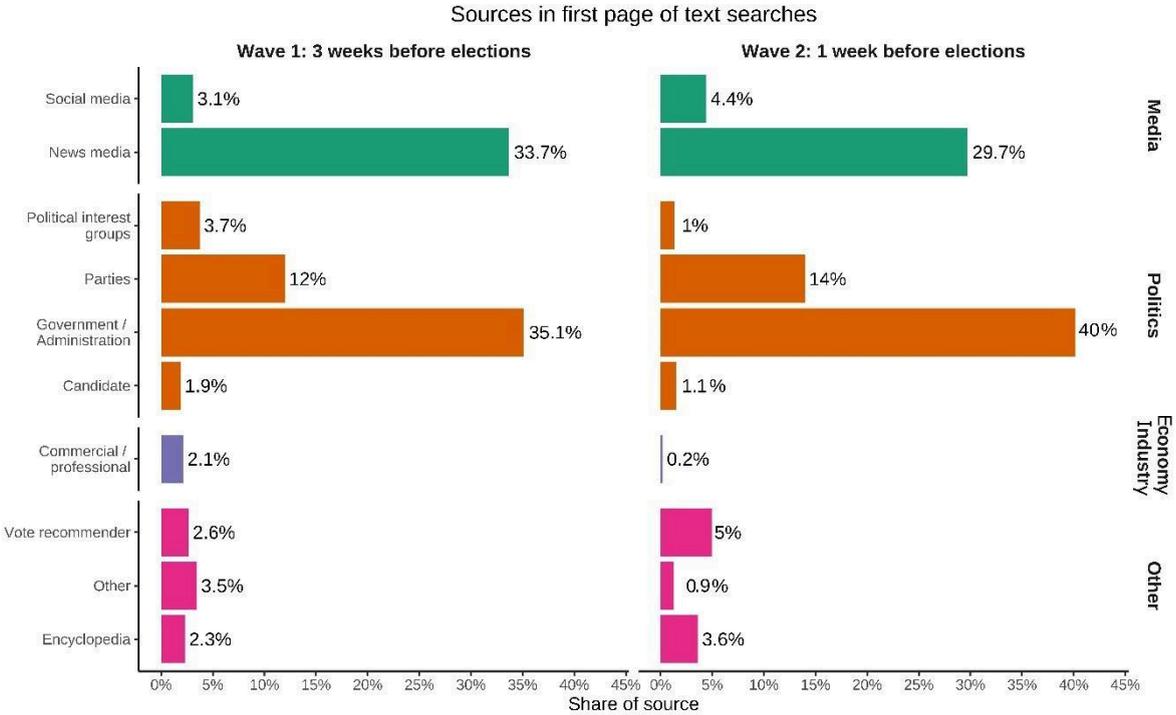

**Figure S1**. Descriptive overview of source types in text search results for both waves of data collection.

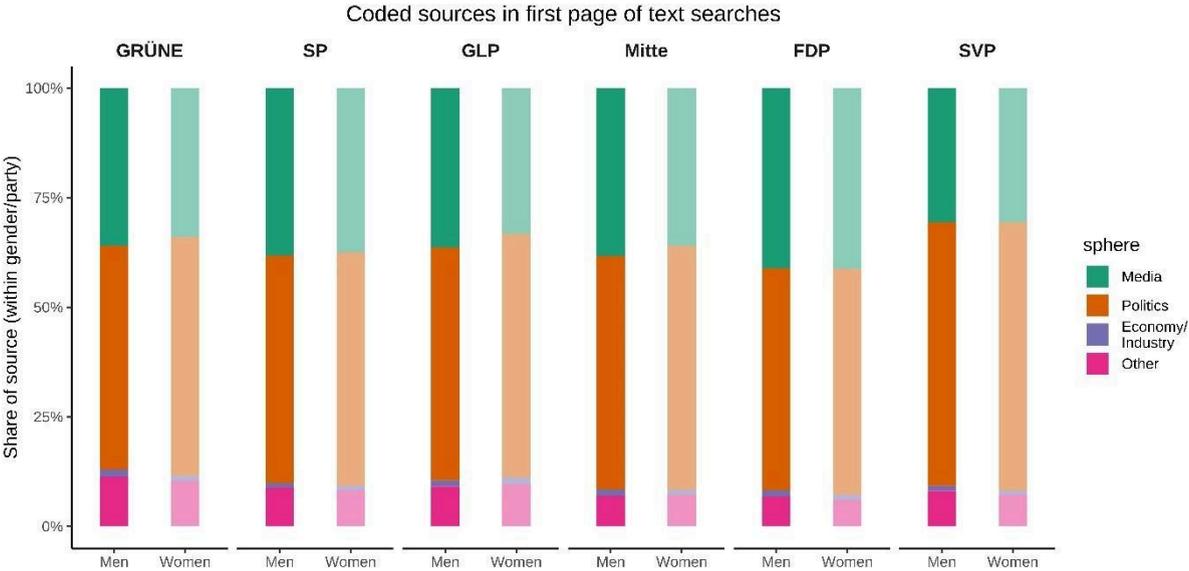

**Figure S2**. Comparison of source type categories by gender.

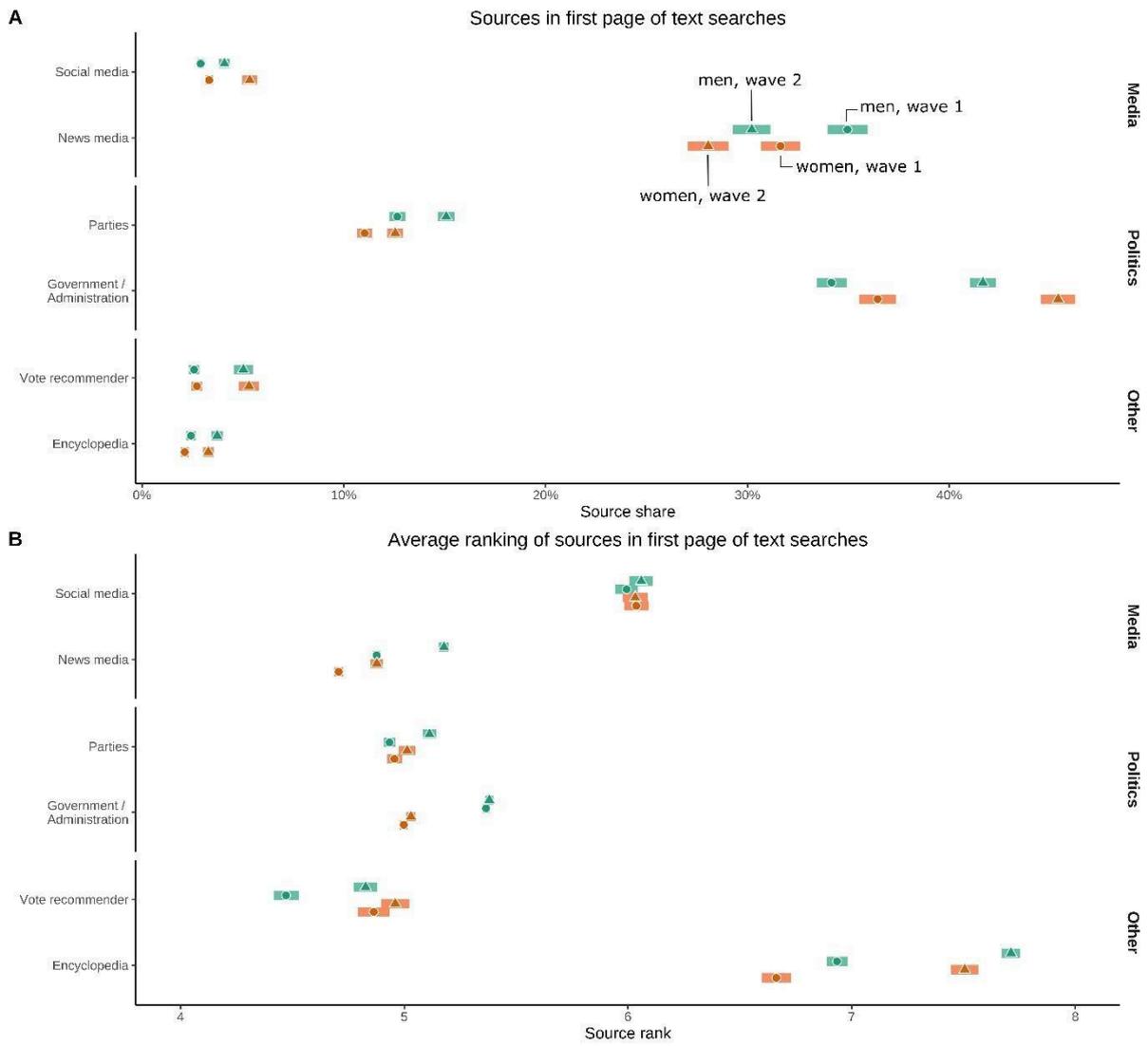

**Figure S3**. Predicting share (Panel A) and ranking (Panel B) of sources in text searches for political candidates running for the 2023 Swiss Federal Elections.